\documentclass[]{interact}

\usepackage{epstopdf}
\usepackage[caption=false]{subfig}

\theoremstyle{plain}

\theoremstyle{definition}

\theoremstyle{remark}

\usepackage{tabularx}
\usepackage{multirow}
\usepackage{hyperref}
\usepackage[table]{xcolor}  

\makeatletter
\renewcommand{\@biblabel}[1]{\quad#1.}
\makeatother

\begin{document}

\articletype{ARTICLE}

\title{Comparative evaluation of the web-based
  contiguous cartogram generation tool go-cart.io}

\author{
\name{Ian~K.~Duncan\textsuperscript{a}\thanks{CONTACT Ian~K.~Duncan. Email: ian.duncan@u.yale-nus.edu.sg} and Michael~T.~Gastner\textsuperscript{b}}
\affil{\textsuperscript{a}Division of Science, Yale-NUS College,
  01-220 Singapore 138527, Singapore\\
  \textsuperscript{b}Infocomm Technology, Singapore Institute of Technology, Singapore
}
}

\maketitle

\begin{abstract}
  Area cartograms are map-based data visualizations in which the area of
each map region is proportional to the data value it represents.  Long
utilized in print media, area cartograms have also become increasingly
popular online, often accompanying news articles and blog posts.
Despite their popularity, there is a dearth of cartogram generation
tools accessible to non-technical users unfamiliar with Geographic
Information Systems software.  Few tools support the generation of
contiguous cartograms (i.e., area cartograms that faithfully represent
the spatial adjacency of neighboring regions).  We thus reviewed
existing contiguous cartogram software and compared two web-based
cartogram tools: fBlog and go-cart.io.  We experimentally evaluated
their usability through a user study comprising cartogram generation
and analysis tasks.  The System Usability Scale was adopted to
quantify how participants perceived the usability of both tools.  We
also collected written feedback from participants to determine the
main challenges faced while using the software.  Participants
generally rated go-cart.io as being more usable than fBlog.  Compared
to fBlog, go-cart.io offers a greater variety of built-in maps and
allows importing data values by file upload.  Still, our results
suggest that even go-cart.io suffers from poor usability because the
graphical user interface is complex and data can only be imported as a
comma-separated-values file.  We also propose changes to go-cart.io
and make general recommendations for web-based cartogram tools to
address these concerns.
\end{abstract}

\section*{Introduction}
\begin{figure}[!ht]
   \includegraphics[width=\linewidth]{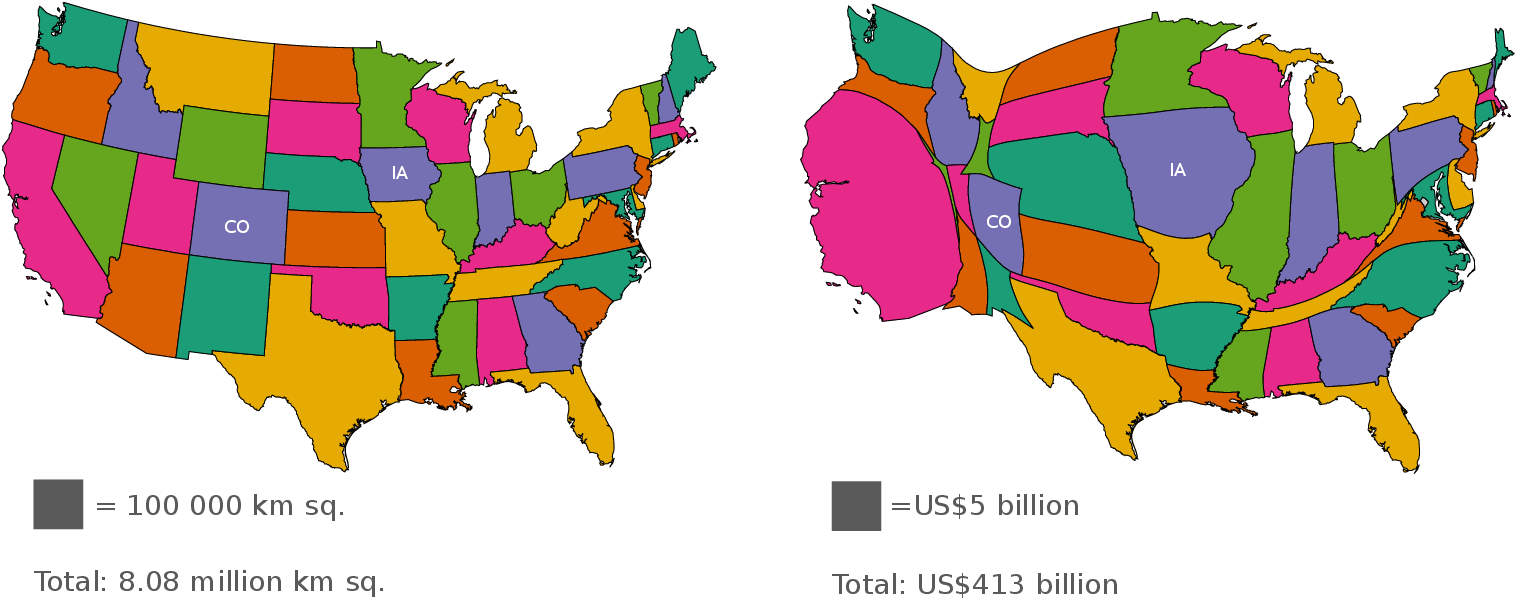}
  \caption{
    {\bf{Example cartogram.}}
    Conventional map (left) and contiguous cartogram (right) of
    2018 agriculture sector output by state in the United
    States~\cite{kassel_state_2021}. Colorado (CO) and Iowa (IA) are labeled in both maps.}
  \label{fig:croptotal-cart}
\end{figure}

Area cartograms are map-based data visualizations in which the area of
each map region (e.g., state or province) is proportional to some
numeric data value (e.g., population or gross domestic product).  An
area cartogram is considered to be contiguous if it preserves
conventional map topology such that neighboring regions on a
conventional (e.g., equal-area) map remain neighbors in the cartogram,
and vice versa.  Consider the contiguous cartogram in
Fig~\ref{fig:croptotal-cart} displaying agriculture sector output
by state in the United States in 2018.  Notice that, while Colorado
(CO) appears larger in area than Iowa (IA) in the conventional map
because Colorado has more land area, Iowa's area appears over three
times larger than that of Colorado in the cartogram.  The proportion
of the cartogram areas reflects that Iowa's agriculture sector output
of US\$$29.5$~billion is more than three times higher than Colorado's
agriculture sector output of
US\$$8.1$~billion~\cite{kassel_state_2021}.  While various types of
non-contiguous cartograms exist, we focus on contiguous cartograms,
which performed well in a previous evaluation that compared different
cartogram types~\cite{nusrat_evaluating_2016}.

First becoming popular in print media in the early 20th
century~\cite{tobler_thirty_2004, hennig_kartogramm_2018}, widespread
adoption of computer technology and the Internet over the last three
decades has created new opportunities for cartograms to be presented
electronically~\cite{ware_using_1998}.  Consequently, cartograms have
become popular in online media as an accompaniment to news articles,
especially those that display statistics, such as election
results~\cite{almukhtar_us_2018, andre_presidential_2020,evershed_building_2013}.

Despite their popularity, there is a dearth of software tools for
generating cartograms that are accessible to non-technical users
unfamiliar with traditional Geographic Information Systems (GIS)
software.  A survey of cartogram generation tools by Markowska and
Korycka-Skorupa~\cite{markowska_evaluation_2015} revealed several web
and desktop applications.  However, presently, many of these tools are
either no longer functional or are designed to be used in conjunction
with a GIS software package, rendering them firmly out of reach of
non-technical users.  Markowska and
Korycka-Skorupa~\cite{markowska_evaluation_2015} do not mention the fBlog Online
Cartogram Tool, which provides a simple web interface for users wanting to
generate cartograms of the United States and
Europe~\cite{van_den_broek_online_2012}.  While its feature set is
limited, it remains one of the only such easy-to-use software tools
that is in working order.
After the survey by Markowska and
Korycka-Skorupa~\cite{markowska_evaluation_2015}, Viry et~al.~\cite{viry_magrit_2016} developed the online mapping application Magrit, which generates cartograms from preinstalled sample basemaps and provides an attractive graphical user interface.
However, its documentation is only available in French, which was not as widely understood as English in the participant pool available for this study.

To address this lack of an easy-to-use and effective cartogram
generation software, Shi, Duncan, and
Gastner~\cite{tingsheng_span_2019} developed go-cart.io, a web
application aimed at non-technical users that generates cartograms of
a variety of geographies from uploaded data.  Herein, we present the
results of an experiment we carried out to evaluate the usability of
go-cart.io and fBlog, which were, to the best of our knowledge, the
only two functional web-based cartogram generation tools
available with user manuals in English at the time of the experiment.  Participants were required to generate a cartogram of a
provided data set with go-cart.io and fBlog.  The order in which
participants encountered the two cartogram tools was randomized.
Afterwards, participants referred to a figure generated from their
cartogram to answer questions about the data set.  Finally,
participants rated the usability of both cartogram generation tools
based on the widely adopted System Usability Scale and left additional
written feedback. Based on the experimental results, we make
recommendations for the design of future versions of web-based
cartogram generation tools.

\section*{Related work}

\subsection*{Usability of GIS software}

The International Organization for
Standardization~\cite{international_organization_for_standardization_ergonomics_2018}
defines system usability as the combination of three criteria:
effectiveness, efficiency, and satisfaction.  Komarkova, Jakoubek, and
Hub~\cite{komarkova_usability_2009} adopted these three criteria as
their framework to evaluate GIS usability. Despite the increasing
prevalence of geospatial data, Komarkova et~al.~\cite{komarkova_usability_2009} note that traditional GIS software
tools suffer from poor usability.  They are often desktop
applications, requiring users to install and configure them before
use.  Often not designed with usability in
mind~\cite{unrau_usability_2019}, GIS user interfaces are usually
complicated and require extensive training before they can be used.
The reliance of certain GIS tools on proprietary file formats is also
a concern for users who need integration with other software
systems~\cite{komarkova_usability_2009}.  The emergence of web-based
GIS tools presents an opportunity to address some of these concerns
because these tools do not need to be installed and configured.
However, web-based applications must overcome certain additional
challenges.  While it may be acceptable for desktop GIS packages to
require extensive training for users to take advantage of their rich
feature set, web applications must cater to users who are not
tech-savvy and are often in a hurry.

In a survey of GIS usability studies, Unrau and
Kray~\cite{unrau_usability_2019} pinpointed additional areas where GIS
tools possess poor usability.  Over half the tools surveyed were
reported to give poor error messages that did not provide any
indication of how to fix the problem encountered.  Moreover, Unrau and
Kray~\cite{unrau_usability_2019} found that many users failed to
complete study tasks because the interface of the studied GIS tool
provided no visual indication of how to proceed.  These findings echo
the results of a survey of users of desktop GIS software conducted by
Davies and Medyckyj-Scott~\cite{davies_gis_1994}.  Participants in
this survey also complained of ``total nonsense'' error messages and
being ``unsure what to do when they sit in front of a GIS''~\cite{davies_gis_1994}.  This latter point may result from the failure
of many GIS software tools to conform to traditional norms of software
interface layout, preventing users from applying general software
skills to GIS systems.  Davies and
Medyckyj-Scott~\cite{davies_gis_1994} also found that most GIS systems
do not do enough to support novice and infrequent users.  They
suggested that GIS software should provide better documentation and
online help because the presence of these features was associated with
a higher usability rating by survey participants.

\subsection*{Cartogram generation tools}

Existing cartogram generation tools also suffer from some of the
usability challenges discussed above.  Markowska and
Korycka-Skorupa~\cite{markowska_evaluation_2015} conducted an
extensive survey of cartogram generation tools.  Out of the five
studied tools, three supported generating contiguous cartograms.  One of these
tools, MAPresso, relies on obsolete Java applet technology and is no
longer usable~\cite{herzog_developing_2003}.  Another tool, ScapeToad was a standalone desktop application but is no longer available online~\cite{andrieu_scapetoad_2008}.  The third tool,
a cartogram utility for ArcGIS~\cite{souza_cartogram_2015}, remains working but is
inaccessible to non-technical users because it is a plugin for
ArcGIS, a commercial GIS software package requiring a software license and specialized
training to use effectively~\cite{markowska_evaluation_2015}.  While all the above-mentioned
tools aim for a high degree of automation, the recently developed
Windows application Cartogram Studio~\cite{kronenfeld_principles_2021}
focuses on manual construction of cartograms.  On the one hand, this
feature makes Cartogram Studio an excellent didactic tool.  On the
other hand, manual customization of cartograms is
time-consuming~\cite{kronenfeld_manual_2018}, limiting the practical
usability of Cartogram Studio outside an educational setting.

Markowska and Korycka-Skorupa~\cite{markowska_evaluation_2015}
quantify the performance of the generation tools they studied using a
numeric scale that awards points for possessing certain functionality
(e.g., whether the tool supports saving generated cartograms in a
vector image format).  However, they do not explicitly consider
usability in their evaluation.  This omission, combined with the
development of new tools such as go-cart.io, provides an opportunity
for an updated, usability-focused assessment of currently available
cartogram generation tools.

\subsection*{System Usability Scale}

Brooke's~\cite{brooke_sus_1996} 1996 System Usability Scale (SUS)
questionnaire provides a standardized method of quantifying the
usability of software and other systems.  The questionnaire comprises
ten statements alternating between positive and negative sentiment.
Table~\ref{tab:sus-questions} lists all the SUS items.  The
alternating sentiment of the questions ensures that participants
carefully consider each questionnaire item~\cite{davies_gis_1994}.
Respondents indicate their agreement or disagreement with each
statement on a $5$-point Likert scale.  Since its development, the SUS
has become widely adopted and has been evaluated as one of the best
performing surveys at measuring system
usability~\cite{lewis_system_2018}.

Unrau and Kray~\cite{unrau_usability_2019} also pointed out that the
SUS has been adopted by many studies that evaluate GIS software
specifically.  Therefore, we adopted the SUS as the main measure of
cartogram generation tool usability in our experiment.

\begin{table}[!ht]
  \centering
  \caption{
    {\bf The 10 items of the System Usabilty Scale~\cite{brooke_sus_1996}.}
    For each item, participants were asked to indicate their agreement with the
    item on a 5-point Likert scale.}
  {\begin{tabularx}{\textwidth}{rX}
		\toprule
		\textnumero & Item Statement \\
		\midrule
		1. & I think that I would like to use this system frequently. \\
		\rowcolor{gray!15}
		2. & I found the system unnecessarily complex. \\
		3. & I thought the system was easy to use. \\
		\rowcolor{gray!15}
		4. & I think that I would need the support of a technical person to be able to use this system. \\
		5. & I found the various functions in this system were well integrated. 
		\\
		\rowcolor{gray!15}
		6. & I thought there was too much inconsistency in this system. \\
		7. & I would imagine that most people would learn to use this system very quickly. \\
		\rowcolor{gray!15}
		8. & I found the system very cumbersome to use. \\
		9. & I felt very confident using the system. \\		
		\rowcolor{gray!15}
		10. & I needed to learn a lot of things before I could get going with this system. \\
		\bottomrule
	\end{tabularx}}
	\label{tab:sus-questions}
\end{table}

\section*{Overview of go-cart.io}

\begin{figure}[!ht]
     \includegraphics[width=\linewidth]{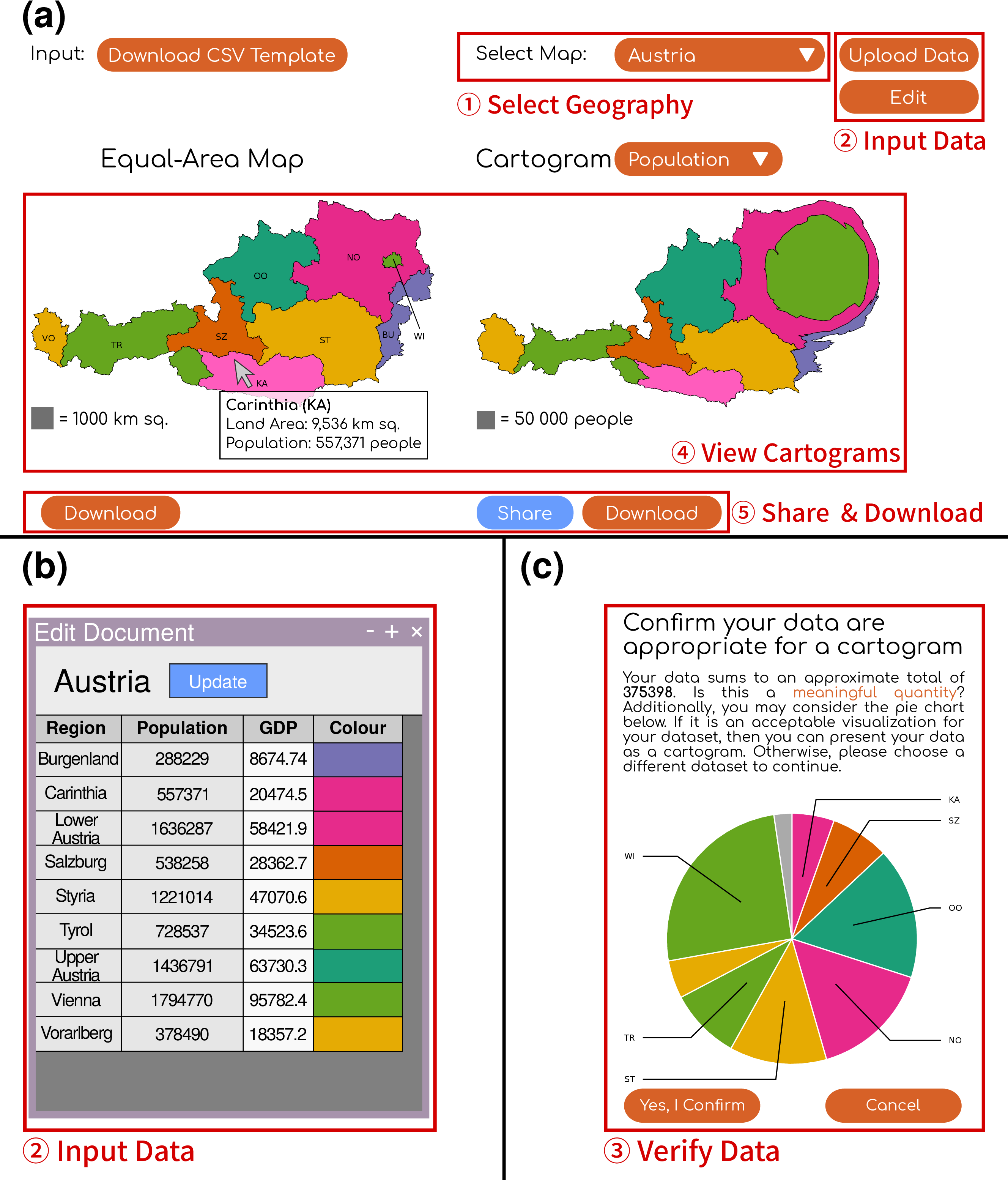}
  \caption{
    {\bf{Screenshot of go-cart.io user interface.}}
    The main interface is presented in panel (a), the pop-up numeric data-entry interface in panel (b), and the pie-chart data verification interface in panel (c).
    Numbers in red circles indicate the sequence of steps required to create a cartogram (see main text for details).}
  \label{fig:gocart-interface}
\end{figure}

The web application go-cart.io allows users to generate cartograms for
a set of selected geographies using their own data.  The generation
tool uses modern web technologies, and runs in any up-to-date web
browser without requiring installation of any additional software.
Table~\ref{table:software-summary} summarizes the features offered by
go-cart.io as compared to existing cartogram generation tools.
Cartograms are generated on a cloud server using the fast flow-based
method developed by Gastner, Seguy, and More~\cite{gastner_fast_2018}.
Fig~\ref{fig:gocart-interface} provides a screenshot of the go-cart.io
interface.  Numbers in the figure highlighting the user interface
elements of go-cart.io correspond to the steps of the instructions
below.  To generate a cartogram, users must:
\begin{enumerate}
\item
    Select the geography for which they want to generate a cartogram
    from the drop-down list at the top-right.  There are currently
    $82$ available geographies on go-cart.io, including countries,
    sub-country divisions (e.g., states and provinces), and
    multinational political entities such as the European Union and
    ASEAN.
    A list of preinstalled maps is available in the online supplemental material on the publisher's website.
\item
    Input numeric data and colors for each region of the selected geography.
	Users may do this in one of two ways:
	
	\begin{itemize}
	\item
	    Downloading a template spreadsheet in comma-separated-values
        (CSV) format by clicking the ``Download CSV Template'' button
        on the top-left.  After entering numeric data as well as
        colors in hex-code format for each region, users may upload
        their edited CSV file by clicking the ``Upload Data'' button
        on the top-right.
	\item
	    Clicking the ``Edit'' button at the top-right and then
        entering the numeric data and color for each region in an
        editing interface that appears in a pop-up window, as shown in
        panel (b) of Fig~\ref{fig:gocart-interface}.
	\end{itemize}

\item
    Confirm that their numeric data are appropriate for a cartogram.
    In most cases, cartograms should only be used for data that add up
    to an interpretable total (e.g., absolute population or gross
    domestic product by region, but not gross domestic product per
    capita~\cite{tingsheng_motivating_2020}).  To aid users,
    go-cart.io displays the users' numeric data in a pie chart, as shown
    in panel (c) of Fig~\ref{fig:gocart-interface}.  If the pie chart
    is an acceptable visualization of the users' numeric data, they may
    proceed to generate the cartogram.  Otherwise, they should select
    different numeric data.
\item
    Once go-cart.io generates a cartogram from the given data, users
    may preview and interact with it.  Following recommendations from
    Dent~\cite{dent_communication_1975}, go-cart.io always displays
    generated cartograms alongside the corresponding conventional map
    and a square-shaped area-to-value legend as anchor stimulus, as
    shown in panel (a) of Fig~\ref{fig:gocart-interface}.
    go-cart.io also provides the following interactive features:
	
	\begin{itemize}
    \item
        \textbf{Infotip}: Hovering the mouse over a region in the
        cartogram or conventional map causes an infotip to appear next
        to the mouse cursor with the region's name, population, land
        area, and numeric data used to generate the cartogram.
	\item
	    \textbf{Linked brushing}: Hovering the mouse over a region in
        the cartogram or conventional map highlights the hovered-over
        region \textit{and} the corresponding region on the other
        map. This feature is implemented by lightening the selected
        color for the region.
	\item
	    \textbf{Map-switching animation}: Using the drop-down menu
        above the cartogram, users may switch between the conventional
        map, population cartogram, and user-generated cartogram.  Each
        time a new map is selected, the currently selected map morphs
        into the newly selected map through a one-second animation.
	\end{itemize}
	
\item
    Users may export or share their generated cartogram by clicking
    the relevant buttons at the bottom of the page.  Cartograms may be
    downloaded in the Scalable Vector Graphics (SVG) format for
    inclusion in a report or presentation as a figure, or in GeoJSON
    format for import into GIS software.  Users may also opt to share
    their generated cartogram on popular social media sites through a
    unique link.
\end{enumerate}

\begin{table}[!ht]
  \centering
  \caption{
    {\bf Summary of automated contiguous cartogram generation software
      tools currently available~\cite{markowska_evaluation_2015,
        van_den_broek_online_2012, tingsheng_span_2019,
        herzog_developing_2003,andrieu_scapetoad_2008,sun_applying_2020}.}}

  {\begin{tabular}{lllll}
	  \toprule
	  
      \multirow{2}{*}{Generation Tool}
      & \multirow{2}{*}{Application Type}
      & Geographic
      & Numeric
      & Interactive\\
      & & Data Source & Data Source
      & Analysis Tools\\
	  
	  \midrule
	  
	  go-cart.io
	  & Web application
	  & $82$ available
	  & CSV upload
	  & Yes\\
	  & & geographies from
	  & or manual &\\
	  & & drop-down menu
	  & entry
	  &\\[0.25em]
	  
	  \rowcolor{gray!15} & & & &\\[-0.75em]
	  \rowcolor{gray!15}
	  Cartogram Utility
	  & ArcGIS plugin
	  & ESRI shapefile
	  & ESRI shapefile
	  & Yes\\
	  \rowcolor{gray!15}
	  for ArcGIS
	  & & & or spreadsheet
	  &\\[0.25em]
	  
	  \\[-0.75em]
	  fBlog
	  & Web application
	  & $2$ available
	  & Manual entry
	  & No\\
	  & & geographies\\[0.25em]
	  
	  \rowcolor{gray!15} & & & & \\[-0.75em]
	  \rowcolor{gray!15}
	  MAPresso &
	  Web application &
	  ESRI shapefile &
	  ESRI shapefile &
	  Yes\\
	  \rowcolor{gray!15}
	  & (obsolete) & & &\\[0.25em]

	  \\[-0.75em]
	  ScapeToad
	  & Desktop application
	  & ESRI shapefile
	  & ESRI shapefile
	  & No\\
	  & \\[0.25em]
	  
	  \rowcolor{gray!15} & & & & \\[-0.75em]
	  \rowcolor{gray!15}
	  F4Carto
	  & Desktop application
	  & ESRI shapefile
	  & ESRI shapefile
	  & No\\
	  \rowcolor{gray!15}
	  &(Windows only) & & &\\[0.25em]
	  \bottomrule
  \end{tabular}}
  \label{table:software-summary}
\end{table}

\section*{Methods}

\subsection*{Cartogram software}

When evaluating the usability of a software system,
Lewis~\cite{lewis_system_2018} recommends performing norms-based
evaluations, where software system usability is evaluated against
standards for usability generally applicable to software systems, and
competitive evaluations.  Norms-based evaluations are insufficient by
themselves because they may be unduly harsh or lenient based on the
category of software being considered.  To conduct a competitive
evaluation of go-cart.io's usability, we performed a survey of similar
cartogram generation tools.  At the time of the experiment, the fBlog Online Cartogram
Tool was the only such tool that is web-based and does not require
users to download any software program or browser plugin.

\begin{figure}[!ht]
     \includegraphics[width=\linewidth]{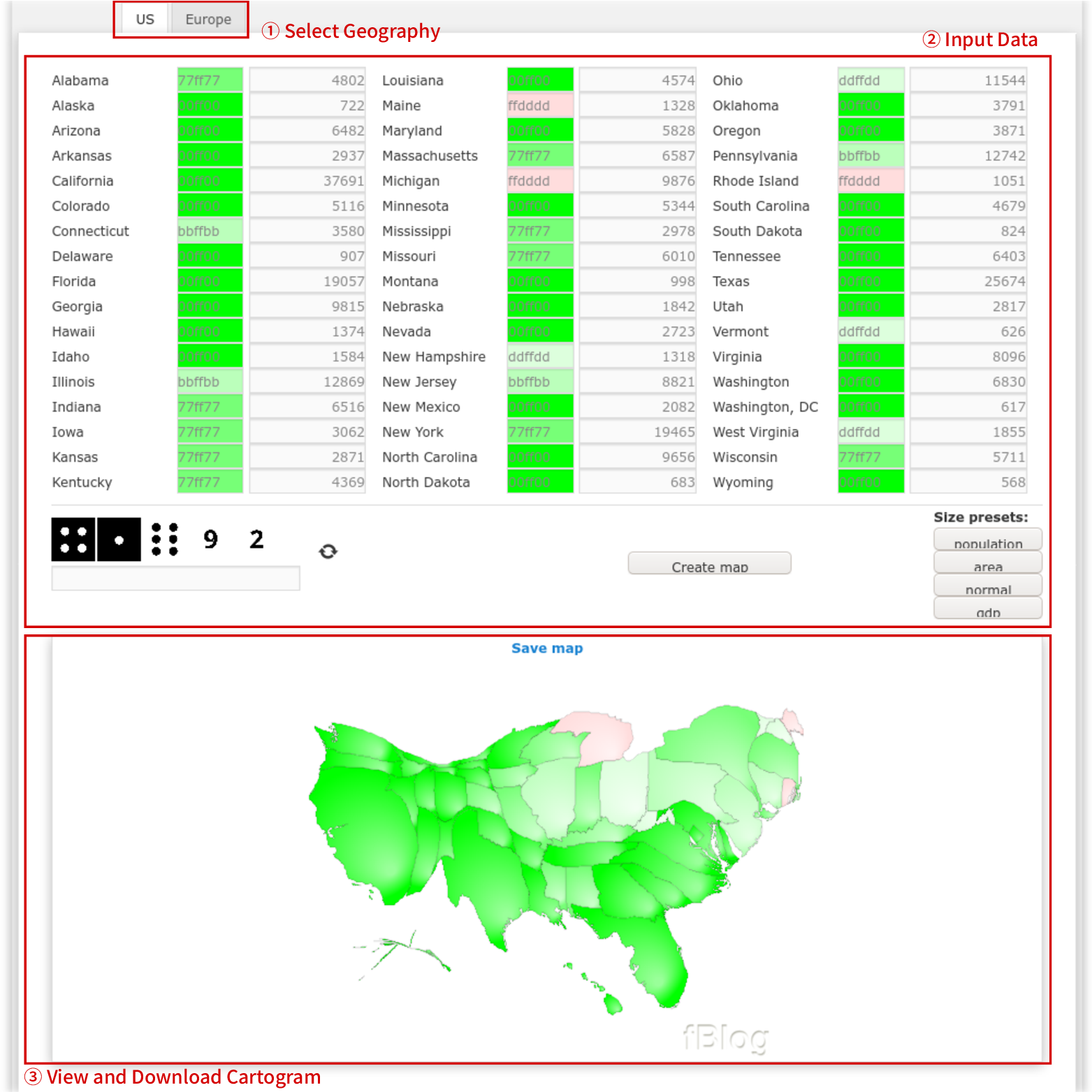}
\caption{ {\bf Screenshot of the fBlog Online Cartogram Tool user interface.}
Important interface elements are highlighted according to their functionality. Numbers in red circles indicate the sequence of steps required to create a cartogram (see main text for details).
Reprinted from~\cite{van_den_broek_online_2012} under a CC BY license, with permission from Korneel van den Broek, original copyright 2012.}
\label{fig:fblog-interface}
\end{figure}

Fig~\ref{fig:fblog-interface} shows a screenshot of the fBlog user
interface, which has a more linear layout than go-cart.io.  To
generate a cartogram, users must:
\begin{enumerate}
\item
    Select whether they would like to generate a cartogram of the
    United States or Europe.
\item
    Input numeric data and colors for each region of the selected
    geography in the text boxes on the page.  Alternatively, users may
    choose from a few preset numeric data sets, including population
    and gross domestic product, by clicking the appropriate button at
    the bottom-right of the numeric input section.
\item
    Fill in the captcha and click ``Create Map''.  The cartogram image
    may be previewed in the browser and downloaded in Portable Network
    Graphics (PNG) format; however, no interactive analysis tools are
    provided.
\end{enumerate}

\subsection*{Tasks}

The experiment comprised cartogram generation and analysis tasks.
During the generation tasks, participants were instructed to generate
a cartogram of a provided data set using one of the two generation
tools (i.e., go-cart.io or fBlog).  Data sets provided for both tools
were for the United States because this is one of two maps available
on both generation tools.  A different data set was used for each
generation tool to avoid a learning effect upon subsequent generation
and analysis tasks.  Both data sets involve agricultural data by
state. To ensure similar difficulty of the analysis tasks, both datasets
had a similar order of magnitude (between $10^8$ and $10^{10}$ USD) and were strongly 
correlated ($r=0.947$).
During the go-cart.io generation task, participants generated
a cartogram of 2018 agricultural output by state~\cite{kassel_state_2021}, while for the fBlog generation task
participants generated a cartogram of 2017 crop sales by state~\cite{united_states_department_of_agriculture_usdanass_2021}.  The
usage of similar data sets helped equalize the difficulty of the
subsequent analysis tasks.  Additionally, we anticipated that most
participants would be unfamiliar with these data sets, thereby
reducing the likelihood of them relying on their own knowledge of the
data sets to complete the analysis tasks.

Participants could not ask the experiment supervisor for help during
generation tasks, but they were allowed to reference a written
tutorial provided by each generation tool on its website.  While the
fBlog tutorial focuses on how to choose good data and colors for a
cartogram, the go-cart.io tutorial provides step-by-step instructions
for generating a cartogram once you already have a data set.
Screenshots of both tutorials are available as online supplemental
material on the publisher's website.  If participants could not
complete a generation task, they were allowed to skip it and proceed
to the tasks for the next generation tool.

Upon completing a generation task, participants were presented with
the cartogram they generated and a correct reference cartogram.
Participants were asked to compare the two by eye.  If they found the
two were \textit{not} identical, participants could reattempt the
generation task to correct their mistakes.  Otherwise, they proceeded
to a set of analysis tasks.

\begin{figure}[!ht]
   \includegraphics[width=\linewidth]{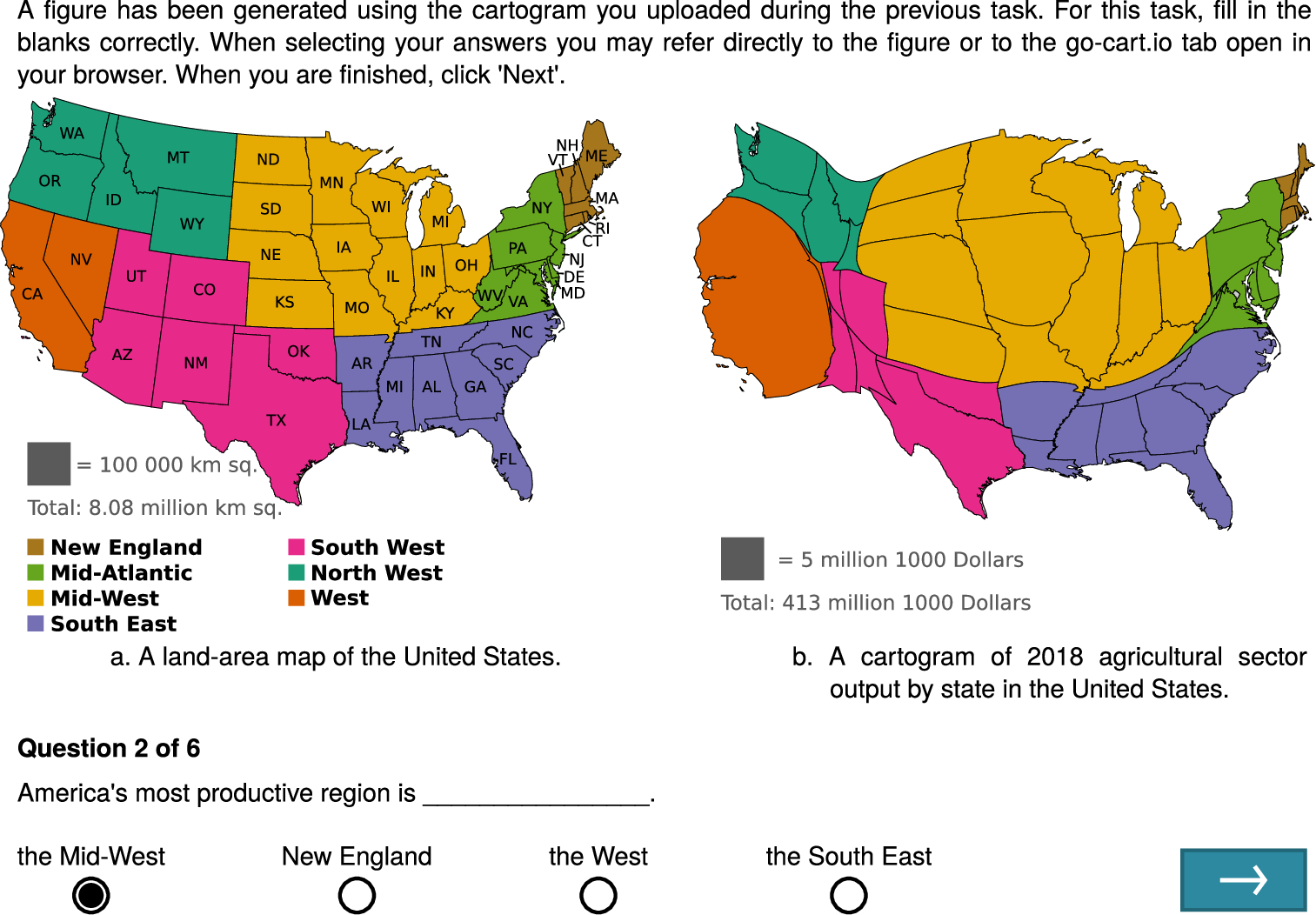}
  \caption{
    {\bf Example cartogram analysis task for go-cart.io.}
    The example task contains a figure created from a correctly
    completed generation task. The figure, question, and answer
    options appeared on the same screen. }
  \label{fig:analysis-task}
\end{figure}

Analysis tasks were designed to simulate how cartograms are used as a
visual aid in reports and presentations.  For each analysis task, a
static figure was generated from the participant's cartogram that
resembled a figure in a report.  Fig~\ref{fig:analysis-task} depicts
an example cartogram analysis task and generated figure.  Using the
figure or any interactive analysis features provided by the generation
tool, participants answered one multiple choice question about the
data set.  The questions in the analysis tasks were loosely inspired
by the cartogram task taxonomy adopted by
Nusrat, Alam, and Kobourov~\cite{nusrat_evaluating_2016}.  All analysis tasks
given to participants during the experiment are available in the
supplemental online material to this article.

\subsection*{Participants}

We recruited $35$ participants to complete the experiment from August 11th to December 18th, 2020.  All
participants were university students and staff.
A departmental ethics review committee at Yale-NUS College approved this research (case number 00016352), and the Institutional Review Board at the National University of Singapore issued a statement of concurrence for the study.
Participants were recruited to the study through the college's online research participation system.
All participants provided informed consent in written form.
The majority of participants reported being at least somewhat familiar
with computer graphics ($20$ participants), spreadsheet software ($28$
participants), and cartograms ($20$ participants).  Half of the
participants ($17$) also indicated that they would at least generally
look up unfamiliar locations on a map.  The participants' age ranged
from $18$ to $28$ (mean $20.8$; standard deviation $1.86$), and their
gender was evenly distributed ($17$ female; $16$ male; $2$ other).
All participants received the equivalent of US\$$7.40$ in local
currency or one hour of research credit for a college course as
compensation for participation.

We administered the Ishihara color blindness test to all participants
because completing the analysis tasks required participants to
distinguish between map regions by color.  Three participants made at
least one error during the test.  However, the responses of these
participants did not differ significantly from the others; thus, we
included them in the data analysis.

We acknowledge that the exclusive participation of university students and staff may primarily represent a demographic of digital natives with higher education, potentially introducing bias to the results. However, the tasks in our study do not require specialized computer skills and are accessible to adolescents and adults with normal vision, including those who wear eyeglasses or contact lenses. Therefore, we believe that our results are broadly applicable to the typical user group of cartogram generation software.

\subsection*{Procedure}

Participants completed the experiment remotely over Zoom using a
single screen.  Each participant's screen was recorded during the
session so that their interactions with the generation tools could be
analyzed later in more detail.  We used Qualtrics XM to display the
experiment tasks and collect participants' answers.  The experiment
comprised four parts:

\begin{enumerate}
\item
  \textbf{Introduction:}
    Participants first watched a short introductory video giving a
    brief overview of cartograms and a description of experiment
    tasks.  They could ask the experiment supervisor for clarification
    at any time.  The video is included in the supplemental online
    material to this article.
\item
  \textbf{Preliminary questions:}
  Participants answered demographic questions about age, gender, and
  level of education.  Then, they indicated their familiarity with
  computer graphics, maps, cartograms, and spreadsheet software on a
  $5$-point Likert scale.  Finally, participants completed the
  Ishihara color blindness test.
\item
  \textbf{Cartogram tasks:}
  Participants completed one cartogram generation task and six
  analysis tasks for each generation tool.  After each set of tasks,
  participants indicated the extent to which they relied on the
  tutorial provided by the generation tool.  If they did indicate that
  they relied on this tutorial, they were also required to indicate
  the helpfulness of the tutorial on a $5$-point Likert scale.
  Finally, participants indicated how much they relied on the
  following sources when completing the analysis tasks:
	
  \begin{itemize}
  \item
	The numbers in the data set table.
  \item
	The cartogram they generated.
  \item
	The interactive analysis tools provided by the generation tool.
  \item
	Their own knowledge of the data set.
  \end{itemize}
	
\item
  \textbf{Usability survey:}
  Participants completed an SUS questionnaire for go-cart.io and
  fBlog.  They then left written, free-form feedback about their
  experience with both web tools.
\end{enumerate}

\subsection*{Design}

We adopted a within-subject experiment design.
Participants completed one trial for each generation tool.  The order
in which participants encountered the tools was treated as a blocking
factor: $18$ participants completed the go-cart.io trial before the
fBlog trial, and $17$ participants completed the fBlog trial before
the go-cart.io trial.  Each trial consisted of one generation task
followed by six analysis tasks.

\subsection*{Hypotheses}

Prior to the experiment, we anticipated that features of go-cart.io
and fBlog would impact how participants rated their relative
usability.  Our hypotheses were as follows:

\subsubsection*{Numeric data input}

While go-cart.io allows users to quickly fill out and upload a CSV
spreadsheet to input numeric data, fBlog requires users to enter data
for each map region manually in its interface.

\begin{itemize}
\item
  \textbf{H1}:
  Participants would find go-cart.io more usable than fBlog because
  of go-cart.io's option to upload numeric data as a CSV file.
\end{itemize}

\subsubsection*{Interactive analysis tools}

While go-cart.io provides interactive analysis tools for generated
cartograms (e.g., infotips), fBlog provides none.
Duncan et~al.~\cite{duncan_task-based_2021} conducted an evaluation of
the interactive analysis tools in go-cart.io and found them to improve
performance during cartogram reading tasks.

\begin{itemize}
\item
    \textbf{H2}:
    Participants would find go-cart.io more usable than fBlog because
    go-cart.io's interactive analysis tools will aid them in
    completing the cartogram analysis tasks.
\end{itemize}

\subsubsection*{User interface layout}

While fBlog has a linear top-to-bottom user interface layout,
go-cart.io provides no clear indication of where to start and how to
proceed with cartogram generation.

\begin{itemize}
\item
  \textbf{H3}:
  Participants would find fBlog more usable than go-cart.io because
  its user interface layout clearly indicates how users should
  proceed.
\end{itemize}

\subsection*{Data analysis}

For analyzing generation task times, we excluded all attempts that resulted in incorrect cartograms (i.e., at least one area or one color differed from the desired output).
Because the generation task times, conditioned on generating correct cartograms, were not normally distributed, we
used the non-parametric paired Wilcoxon ranked-sign test to compare
generation task times between generation tools.  For this and other
tests, we considered $p$-values as significant if they are less than
$0.05$, and we also computed confidence intervals.
We adopted the method developed by
Bauer~\cite{bauer_constructing_1972} to estimate the $95$\% confidence
interval of the pseudomedian difference in generation task times
between the two tools.

For rating participants' accuracy on the generation tasks, we analyzed
mistakes made by entering the wrong color and wrong numeric data for
each region separately as well as the total number of errors of all
types made by each participant.  Similarly, for the analysis tasks, we
analyzed the error rates for each task as well as the number of
analysis tasks performed correctly for each generation tool.  We
considered an analysis task to be performed correctly only if all
parts of the task were completed correctly.  A permutation test with
$10\,000$ random simulations was used to determine the effect of
generation task accuracy on the number of correct analysis tasks for
each generation tool.

For the SUS, we computed the score for each participant using the
standard methodology presented by Brooke~\cite{brooke_sus_1996}.  We
used Cronbach's alpha with an acceptability range of
$0.7<\alpha<0.95$, as recommended by Lewis~\cite{lewis_system_2018} to
evaluate the internal consistency of the SUS questions as applied to
this experiment.  Because SUS scores were approximately normally
distributed, we used a paired $t$-test to evaluate the effect of
generation tool on mean SUS score.  We also used Welch's unequal
variances $t$-test to evaluate the effect of generation task
completion on mean SUS score for each tool.

Finally, we analyzed whether the hypotheses we presented in the
previous section are supported.  For H1, we used Spearman's
$\rho$, which evaluates how well a monotonic function describes the
relationship between two numeric variables, to detect if there is a
positive correlation between participants' indicated familiarity with
spreadsheet software and SUS score for go-cart.io.  We adopted the
bootstrapping method presented by
Bishara and Hittner~\cite{bishara_confidence_2017} to compute $95$\%
confidence intervals for $\rho$.  For H2, we also used
Spearman's $\rho$ to see if there was a positive correlation between
participants' indicated reliance on go-cart.io's interface to complete
the go-cart.io analysis tasks and SUS score for go-cart.io.  For
H3, we considered participants' reliance on the tutorial
provided by the generation tool during the tool's generation task as
an indication that the tool's interface is unclear.  We used a
permutation test with $10\,000$ simulations to determine if reliance
on the tutorial was significantly higher for go-cart.io than for
fBlog.

\section*{Results}

\subsection*{Generation tasks}

\subsubsection*{Completion}

Most participants completed both generation tasks.  $32$ out of $35$
participants finished the generation task with fBlog, while $26$
participants finished the generation task with go-cart.io; $23$ (i.e.,
$65$\% of the participants) completed both generation tasks.  One
participant was not able to complete the go-cart.io generation task
due to a technical error with the go-cart.io web application, and
another participant mistakenly used fBlog to complete the go-cart.io
generation task.  Data from these two participants have been excluded
from the remaining analysis.

\subsubsection*{Duration}

\begin{figure}[!ht]
   \includegraphics[width=\linewidth]{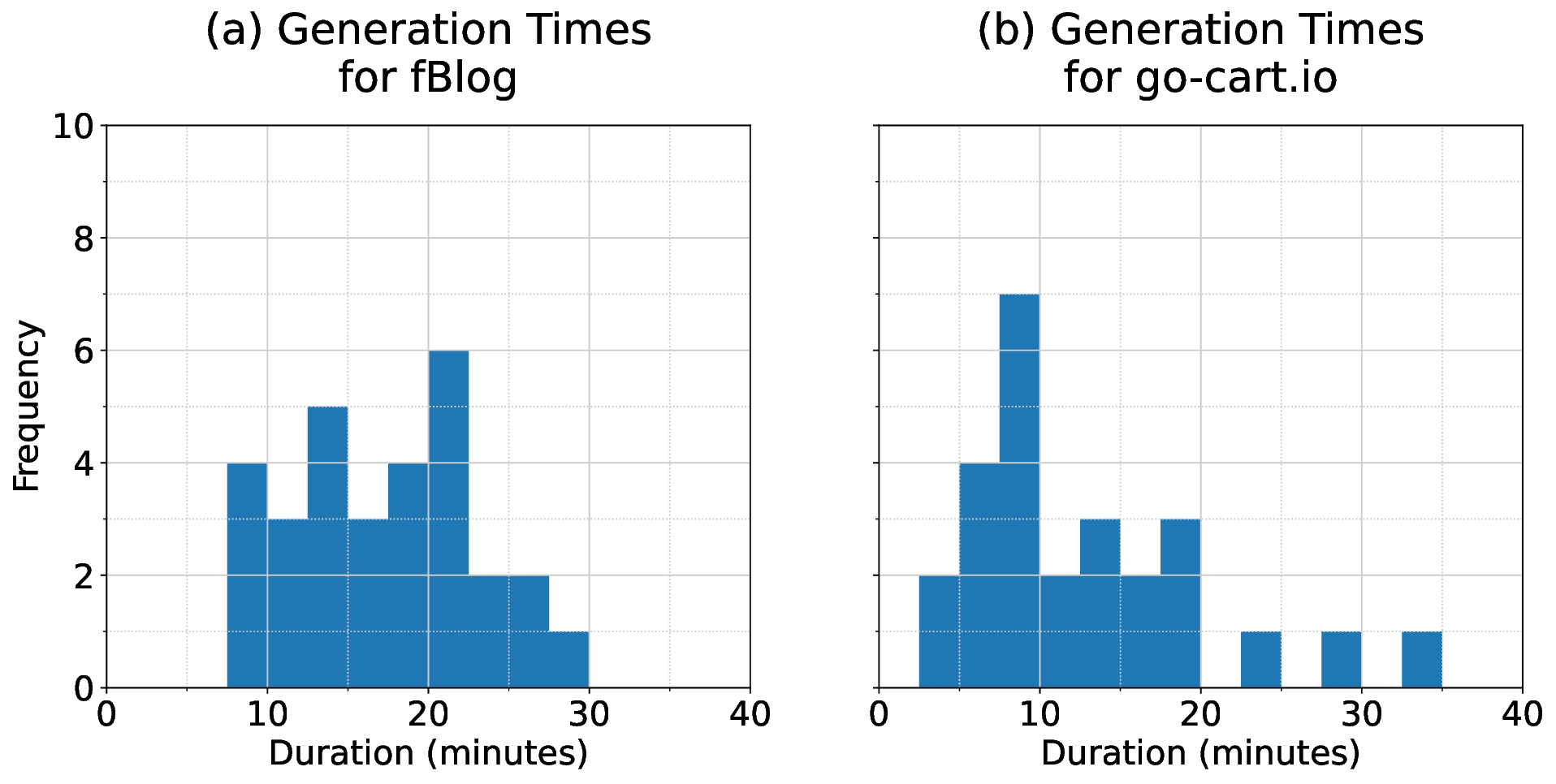}
  \caption{
    {\bf Generation task duration by generation tool. }
    Distribution of duration in minutes taken by participants to
    complete both generation tasks.}
  \label{fig:duration_distribution}
\end{figure}

Median generation task duration for fBlog, conditioned on generating the correct cartogram, was $17.2$ minutes versus
$10.2$ minutes for go-cart.io ($95$\% confidence interval for
pseudomedian difference in minutes: $[0.442, 8.77]$).
Fig~\ref{fig:duration_distribution} shows the distribution of
generation task duration for both tools.  The distribution of duration
for fBlog was approximately uniform, with a minimum time of $7.62$
minutes and maximum time of $28.12$ minutes.  By contrast, the
distribution of duration for go-cart.io was right-skewed, with a
minimum time of $3.70$ minutes and maximum time of $34.45$ minutes.
The difference in pseudomedian duration between the two tools was
statistically significant ($p=0.038$).

\subsubsection*{Accuracy}

Most participants who finished the generation tasks completed these
tasks accurately.  $69.2$\% of the participants who completed the
go-cart.io generation task completed it with perfect accuracy, while
this number was $50$\% for the fBlog generation task.
Table~\ref{table:generation_accuracy} provides a breakdown of the
accuracy and error rates for both fBlog and go-cart.io generation
tasks.  Overall, participants' accuracy in entering state areas was
high across both tasks.  $92.3$\% of the participants completing the
go-cart.io generation task entered all $49$ region areas correctly,
while $81.2$\% of the participants completing the fBlog generation
task entered all $51$ region areas correctly.\footnote{fBlog includes
  Alaska and Hawaii in the map of the United States, whereas the maps
  on go-cart.io only contain the states in the conterminous United
  States and Washington, D.C.}  For both tasks, accuracy of region
colors was lower.  $69.2$\% of the participants who completed the
go-cart.io generation task entered all region colors correctly, while
only $56.3$\% did so for the fBlog generation task.

\begin{table}[!ht]
  \centering
  \caption{
    {\bf Generation task accuracy by generation tool. }
    Accuracy and error rates for region areas and colors in the (a)
    fBlog and (b) go-cart.io generation tasks.  ``With Error'' refers
    to the proportion of participants making at least one error in
    entering the area or color of a map region.}

  {
  	\renewcommand{\arraystretch}{1.2}
	\centering
	\begin{tabular}{@{}cc cc@{}}
		\multicolumn{4}{c}{\textbf{(a) fBlog}} \\
		\cline{2-4}
		\multicolumn{1}{c}{} &\multicolumn{1}{c}{} &\multicolumn{2}{c}{Areas} 
		\\ 
		\cmidrule(lr){3-4}
		\multicolumn{1}{c}{} & 
		\multicolumn{1}{c}{} & 
		\multicolumn{1}{c}{Accurate} & 
		\multicolumn{1}{c}{With Error} \\
		\cline{2-4}
		\multirow[c]{2}{*}{\rotatebox[origin=tr]{90}{Colors}}
		& Accurate  & $0.500$ & $0.063$   \\[1.2ex]
		& With Error  & $0.313$   & $0.125$ \\ 
		\cline{2-4}
		& & & \\
		\multicolumn{4}{c}{\textbf{(b) go-cart.io}} \\
		\cline{2-4}
		\multicolumn{1}{c}{} &\multicolumn{1}{c}{} &\multicolumn{2}{c}{Areas} 
		\\ 
		\cmidrule(lr){3-4}
		\multicolumn{1}{c}{} & 
		\multicolumn{1}{c}{} & 
		\multicolumn{1}{c}{Accurate} & 
		\multicolumn{1}{c}{With Error} \\
		\cline{2-4}
		\multirow[c]{2}{*}{\rotatebox[origin=tr]{90}{Colors}}
		& Accurate  & $0.692$ & $0.000$   \\[1.2ex]
		& With Error  & $0.231$   & $0.077$ \\ 
		\cline{2-4}
  \end{tabular}}
  \label{table:generation_accuracy}
\end{table}

\begin{figure}[!ht]
   \includegraphics[width=\linewidth]{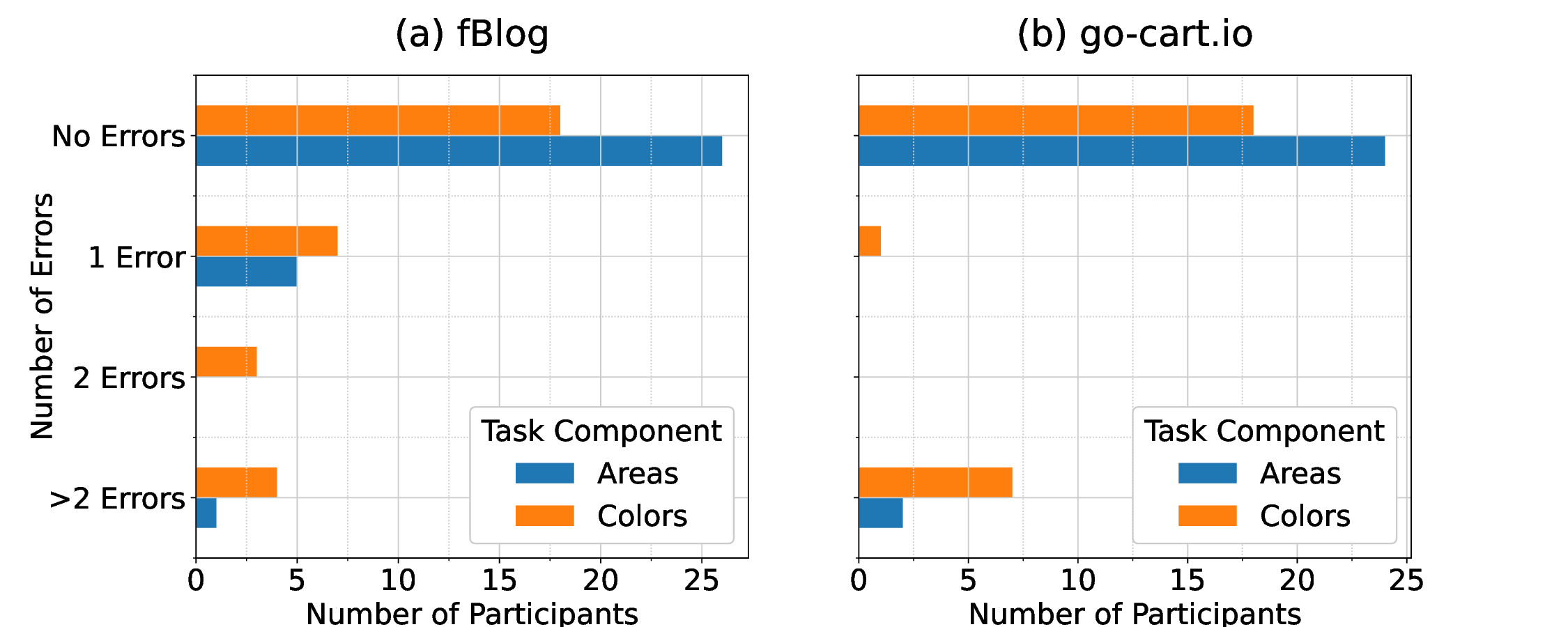}
  \caption{
    {\bf Generation task errors by generation tool. }
    Distribution of the number of area and color errors made by
    participants during the (a) fBlog generation task and (b)
    go-cart.io generation task.}
  \label{fig:generation_errors_distribution}
\end{figure}

Among participants who made errors during the fBlog generation task,
the number of errors was usually small.
Fig~\ref{fig:generation_errors_distribution} shows the distribution
of the number of area and color errors for both generation tasks.  For
the fBlog generation task, $5$ out of the $6$ participants who made an
area error made only one such error.  Similarly, $10$ out of the $14$
participants who made a color error on the fBlog task made at most two
such errors.

However, the opposite is true for the go-cart.io generation task.
Both participants who made an area error made at least three such
errors, and $7$ out of the $8$ participants who made a color error
made at least three such errors.

\subsubsection*{Reliance on and helpfulness of tutorial}

\begin{figure}[!ht]
   \includegraphics[width=\linewidth]{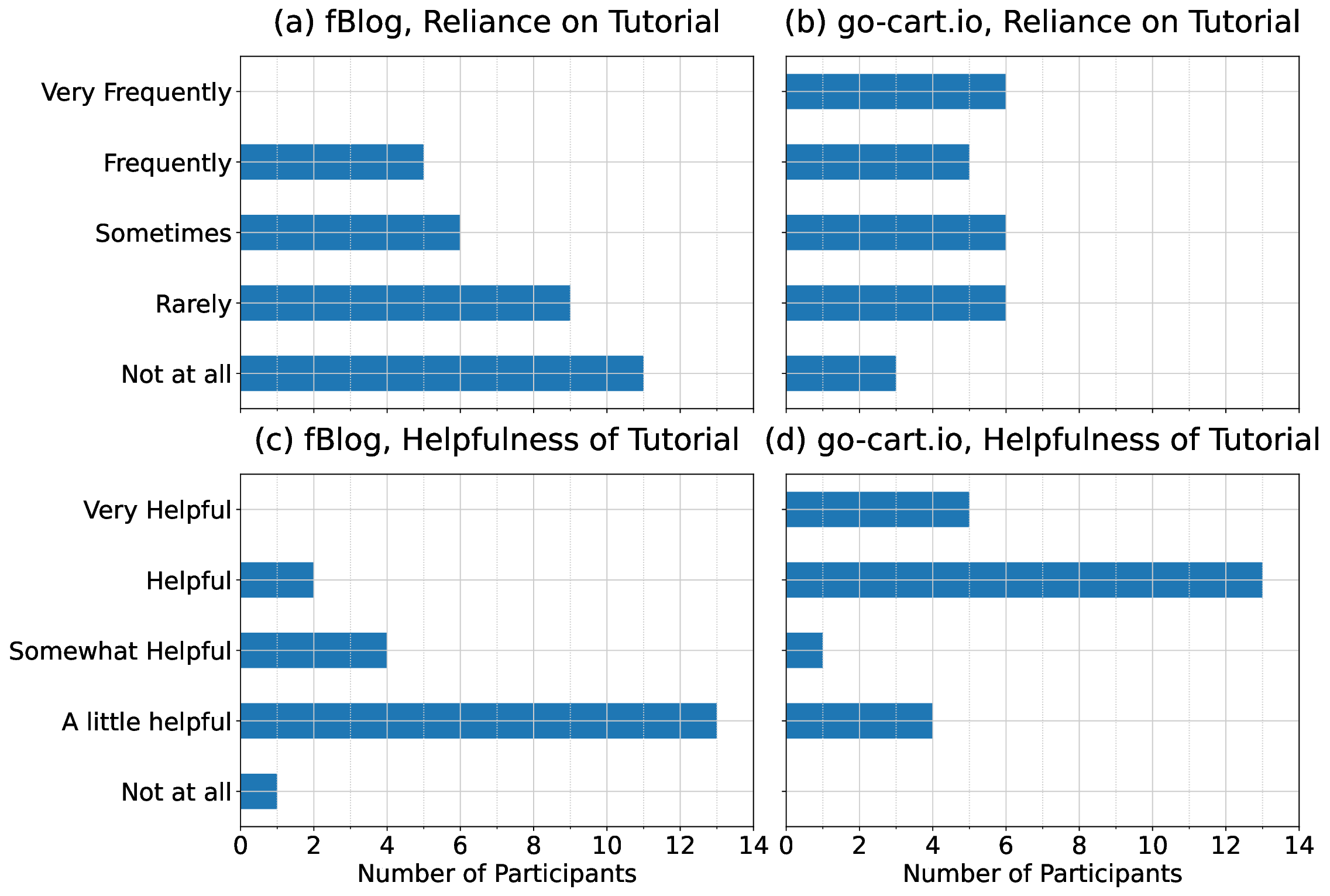}
  \caption{
    {\bf Self-reported reliance and helpfulness of generation tool tutorials.}
    Panels (a) and (b) show how often participants indicated that they
    relied on the written tutorials provided by fBlog and go-cart.io,
    respectively, during each generation task.  For participants who
    indicated that they relied on the tutorial for a generation tool
    at least ``Rarely'', panels (c) and (d) show how helpful
    participants rated the respective tutorials for fBlog and
    go-cart.io.}
  \label{fig:tutorial}
\end{figure}

Participants indicated lower reliance on the tutorial during the fBlog
generation task than during the go-cart.io generation task.
Fig~\ref{fig:tutorial} provides an overview of participants'
indicated reliance on and helpfulness of the tutorials during both
generation tasks.  Considering participants' responses on an interval
scale from $0$ (``Not at all'') to $4$ (``Very frequently''), mean
reliance on the tutorial was found to be $1.03$ points higher for
go-cart.io than for fBlog ($95$\% confidence interval $[0.491, 1.58$]).
Mean reliance on the tutorial for fBlog was $2.19$, while it was
$1.16$ for fBlog.  This difference is significant ($p=0.002$).

Among participants who relied on a tutorial, the mean helpfulness of
the go-cart.io tutorial was $1.48$ points higher than the mean
helpfulness of the fBlog tutorial ($95$\% confidence interval $[0.962, 1.98]$),
on an interval scale from $0$ (``Not at all'') to $4$ (``Very
helpful'').  Mean helpfulness was $1.35$ for the fBlog tutorial and
$2.83$ for the go-cart.io tutorial.  This difference is also
significant ($p<0.001$).

\subsection*{Analysis tasks}

\subsubsection*{Error rates}

\begin{figure}[!ht]
   \includegraphics[width=\linewidth]{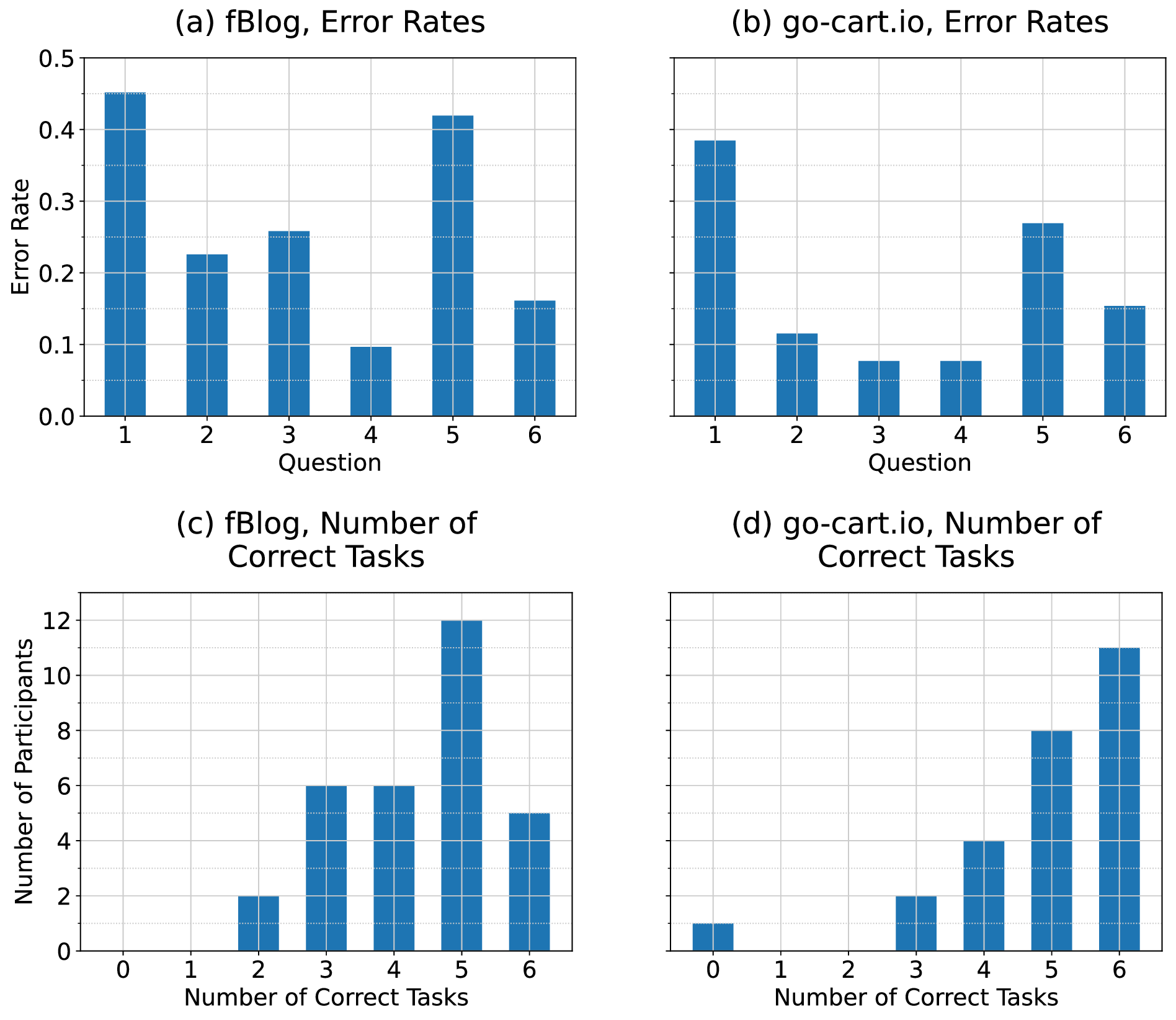}
  \caption{
    {\bf Participants' performance on analysis tasks.}
    Panels (a) and (b) show error rates by analysis task question for
    fBlog and go-cart.io, respectively.  Panels (c) and (d) show the
    distribution of the number of correct tasks for fBlog and
    go-cart.io, respectively.}
  \label{fig:analysis_error_rates}
\end{figure}

Participants found most analysis tasks for both fBlog and go-cart.io
to have moderate difficulty, with some tasks being unexpectedly very
difficult.  Fig~\ref{fig:analysis_error_rates} provides an overview
of participants' performance on the analysis tasks.  The first
analysis task for both generation tools had the highest error rates
($0.452$ for fBlog and $0.385$ for go-cart.io).  This may have been
the partial result of the first analysis task having two parts for
both generation tools (participants had to name the largest and
second-largest region), while all other analysis tasks had only one
part.

\subsubsection*{Effect of generation tool}

For both generation tools, a majority of the participants completed
over half of the analysis tasks correctly ($74.2$\% for fBlog and
$88.5$\% for go-cart.io).

\subsubsection*{Effect of generation task accuracy}

Participants who completed the fBlog generation task without error
completed on average $0.467$ ($95$\% confidence interval $[-0.067, 1.00]$)
more fBlog analysis tasks correctly than those who made at least one
error of any type during the generation task.  However, fBlog
generation task accuracy did not have a significant effect on the
number of fBlog analysis tasks completed correctly ($p=0.145$).

Participants who completed the go-cart.io generation task without
error completed on average $0.611$ ($95$\% confidence interval
$[-0.795, 2.03]$) more go-cart.io analysis tasks correctly than
participants who made at least one generation task error.  The
accuracy of go-cart.io generation tasks did not have a significant
effect either on the number of go-cart.io analysis tasks completed
correctly ($p=0.147$).

\subsubsection*{Participants' indicated methodology}

\begin{figure}[!ht]
   \includegraphics[width=\linewidth]{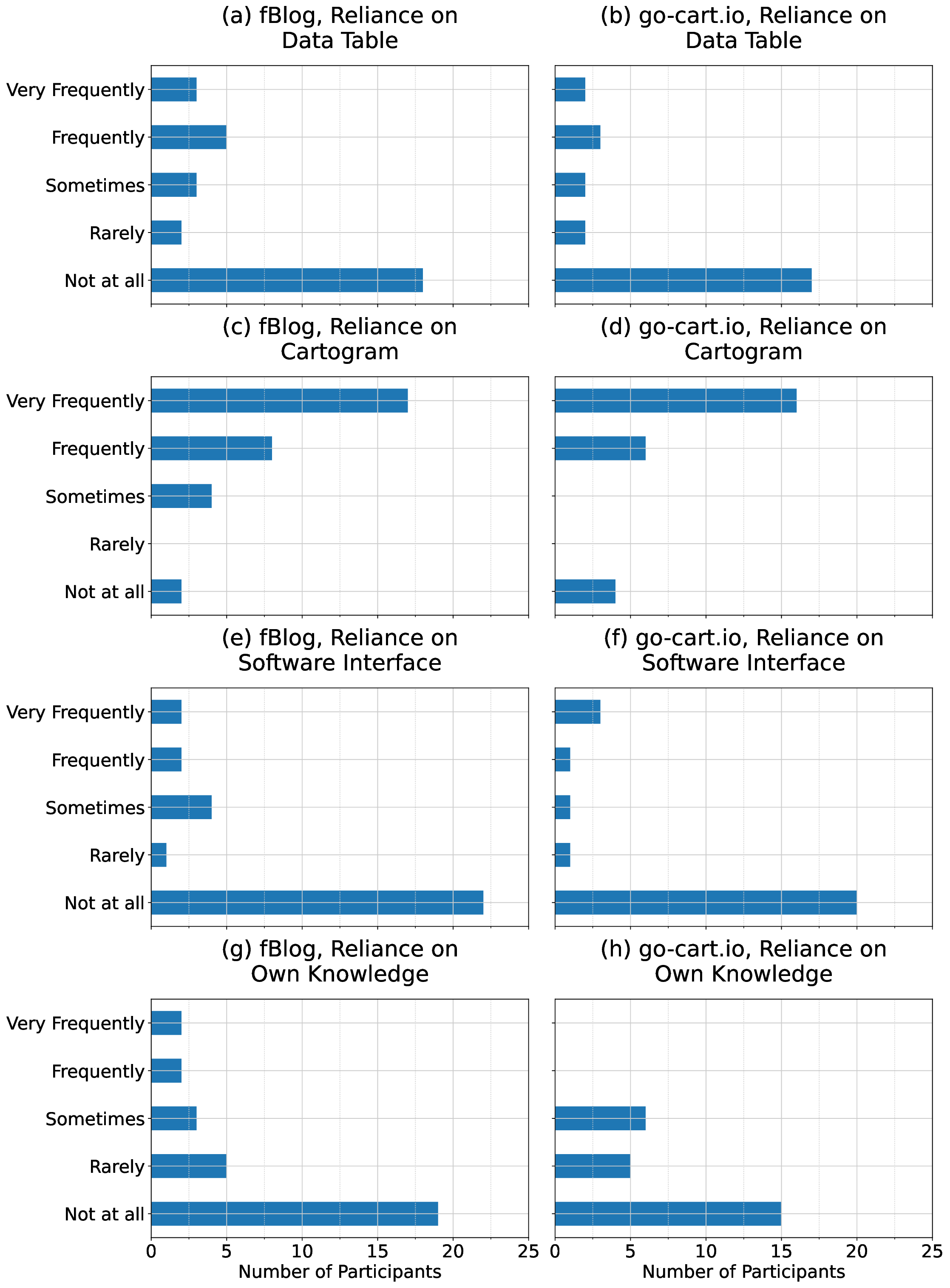}
  \caption{
    {\bf Participants' indicated methodology for completing analysis tasks.}
    Participants' indicated reliance on the data table [panels (a) and
    (b)], generated figure with cartogram [panels(c) and (d)],
    software interface [panels (e) and (f)], and their own knowledge
    [panels (g) and (h)] while completing the analysis tasks for fBlog
    and go-cart.io.}
  \label{fig:howanswer}
\end{figure}

Most participants relied on the figure created from their cartogram
during the generation tasks for both generation tools.
Fig~\ref{fig:howanswer} provides an overview of the sources of
information participants indicated they relied on while completing the
analysis tasks.  $80.6$\% of the participants who completed the fBlog
analysis tasks reported relying on the generated cartogram figure
frequently or very frequently, while $84.6$\% who completed the
go-cart.io analysis tasks indicated the same.

A minority relied on the numbers in the data table or their own
knowledge to complete the analysis tasks.  Only $35.5$\% of those
completing the fBlog analysis tasks reported relying on numbers in the
data table more than rarely, and $22.6$\% relied on their own
knowledge more than rarely.  Similarly, $26.9$\% of the participants
completing the go-cart.io analysis tasks indicated that they relied on
the numbers in the data table more than rarely, and $23.1$\% relied on
their own knowledge more than rarely.

Reported reliance on the generation tool interface, including
interactive analysis tools, was also low.  Only $24.4$\% of those
completing the fBlog analysis tasks reported relying on the fBlog
interface, while $19.9$\% indicated the same for go-cart.io.

\subsection*{System Usability Scale score}

The SUS is a highly reliable measure of perceived usability for both
fBlog ($\alpha=0.860$) and go-cart.io ($\alpha=0.923$).

The mean SUS score was $18.2$ points higher for go-cart.io than for
fBlog ($95$\% confidence interval $[6.78, 29.6]$).  While the mean SUS
score for fBlog was $47.1$ (standard deviation $19.4$), the mean SUS
score for go-cart.io was $65.3$ (standard deviation $23.0$).  The
difference in mean SUS score for fBlog and go-cart.io was significant
($p = 0.003$).

Completion of the generation task for fBlog and go-cart.io was
associated with a higher mean SUS score.  The mean fBlog SUS score for
participants failing to complete the fBlog generation task was $30$;
however, for those who completed this task, the mean SUS score was
$48.0$ ($95$\% confidence interval of difference in means: $[7.93, 28.1]$).
Similarly, while the mean go-cart.io SUS score was $38.6$ for
participants who did not complete the go-cart.io generation task, the
mean SUS score was $72.5$ for those who did complete this task ($95$\%
confidence interval of difference in means $[16.7, 51.1]$).  For both
fBlog ($p=0.003$) and go-cart.io ($p=0.001$), completion of generation
task had a significant effect on mean SUS score.

Participants' indicated familiarity with spreadsheet software was not
significantly correlated with the SUS score for fBlog ($\rho=0.318$,
$95$\% confidence interval $[-0.03, 0.596]$) or go-cart.io
($\rho=0.014$, $95$\% confidence interval $[-0.331, 0.355]$.  Reliance
on the generation tool interface, including any interactive analysis
tool, during the analysis tasks was not significantly correlated with
SUS score for fBlog ($\rho=0.190$, $95$\% confidence interval
$[-0.176, 0.510]$) or go-cart.io ($\rho=0.236$, $95$\% confidence
interval $[-0.166, 0.571]$) either.

\subsection*{Written participant feedback}

\subsubsection*{fBlog}

Participants reported using fBlog to be ``troublesome'' and
``tedious'' due to the manual numeric data-entry method.  They also
indicated that they would have preferred a spreadsheet upload option,
as implemented in go-cart.io, or the ability to copy tabular data
directly into the web interface.

\subsubsection*{go-cart.io}

Several participants indicated that they found the go-cart.io
interface to be aesthetically pleasing.  While a few participants
wrote that the generation tool was easy to use, most indicated that
they faced issues using the spreadsheet upload feature to input
numeric data for the generation task.  Participants complained that
the instructions for formatting and saving the spreadsheet were
unclear.  Some participants were unaware that only CSV spreadsheet
files could be uploaded and expressed frustration that
Microsoft\textsuperscript{\textregistered} Excel spreadsheets could
not be uploaded.  Many of the participants who eventually succeeded
indicated that they relied heavily on go-cart.io's written tutorial
and had to read it carefully.  Participants who gave up trying to use
the spreadsheet upload feature and instead used the pop-up editing
interface [shown in panel (b) of Fig~\ref{fig:gocart-interface}]
complained that entering map region colors was difficult because color
codes could not be pasted into the editing interface.

\subsection*{Hypotheses}

\subsubsection*{H1 is rejected: go-cart.io's numeric data input is not more usable than that of fBlog}

While several participants indicated in their written feedback that
they preferred go-cart.io's spreadsheet upload option to fBlog's
manual input method, most participants struggled to utilize the former
method.  Additionally, we did not find a significant relationship
between participants' familiarity with spreadsheet software and their
perceived usability of go-cart.io.  For these reasons, H1 is
rejected.

\subsubsection*{H2 is rejected: go-cart.io's interactive analysis tools played no significant role for usability}

Few participants made use of the interactive analysis tools provided
by go-cart.io to complete the go-cart.io analysis tasks, and there was
no significant correlation between increasing reliance on go-cart.io's
interactive analysis tools and SUS score.  For these reasons,
H2 is rejected.

\subsubsection*{H3 is partially supported: go-cart.io's user interface layout leads to more frequent referencing of tutorials that that of fBlog}

Although mean SUS score was significantly lower for fBlog than for
go-cart.io, participants relied on the tutorial to complete the
go-cart.io generation task significantly more than during the fBlog
generation task.  This greater reliance implies that more participants
were unable to use go-cart.io without guidance than fBlog.
H3 is thus partially supported.

\section*{Discussion}

A majority of the participants were able to complete the generation
and analysis tasks for go-cart.io with reasonable accuracy.
Participants who did make errors on the go-cart.io generation task
were more likely to have made many more errors than for the fBlog
generation task, where those who made errors generally made only one
or two.  We hypothesize that the differing numeric data and color
entry methods of the two tools account for this difference.  For the
go-cart.io task, participants were able to copy numeric data and
region colors as columns and paste them into a CSV spreadsheet for
upload.  This method of data entry means that likely sources of error,
such as transposition of spreadsheet columns, would lead to many
errors in the generated map.  By contrast, fBlog's manual entry method
caused each typo or incorrect pasting to affect only one region.

The time savings from using the spreadsheet upload feature as compared
to manual entry of numeric and color data for each region were
substantial, accounting for the significant difference in median
generation time between fBlog and go-cart.io.  To help users detect
when they have made errors in the spreadsheet they upload to
go-cart.io, the application shows uploaded numeric data in pie chart
form and asks users to confirm that the data are correct and
appropriate for a cartogram before proceeding to the generation phase.
Panel (c) of Fig~\ref{fig:gocart-interface} shows the pie-chart
display.

Yet, these time savings did not translate into a high usability rating
of go-cart.io by participants.  The mean SUS score of $65.6$ for
go-cart.io is slightly lower than the mean SUS score of $65.7$ for
desktop GIS software tools found by
Davies and Medyckyj-Scott~\cite{davies_gis_1994}.  Like the
respondents to their 1994 survey, participants in this experiment
complained that go-cart.io was unintuitive to use and had poor error
messages.  In the following sections, we make recommendations to
improve the usability of go-cart.io as a web-based GIS tool.  We
believe these recommendations will also be informative for the
development of other web-based cartogram generation tools.

\subsection*{Recommendations}

\subsubsection*{Data entry}

Entering the numeric data and color for each map region during the
go-cart.io generation task proved to be one of the most difficult
tasks for participants during the experiment.  A variety of factors
are responsible for making this task more difficult than anticipated.

First, while go-cart.io only accepts CSV-format spreadsheets for
upload, several participants were unaware of the difference between
spreadsheet formats and erroneously assumed they could upload
Excel-format spreadsheets.

\begin{itemize}
\item 
    \textbf{Recommendation 1:}
    Due to the format's popularity, web-based cartogram generation tools should accept Excel-format spreadsheet files in addition to CSV-format spreadsheet files. Support for additional file formats, such as Open Document Format Spreadsheet (ODS) and JavaScript Object Notation (JSON), is also desirable.
\end{itemize}

\begin{figure}[!ht]
   \includegraphics[width=0.5\linewidth]{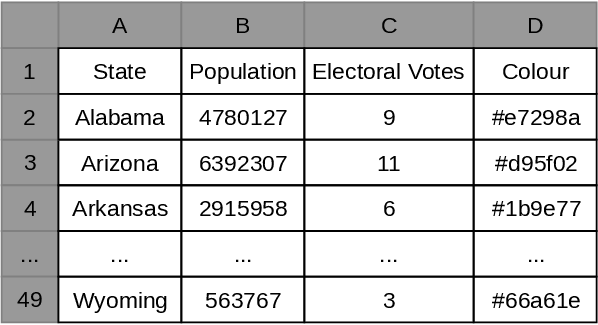}
  \caption{
    {\bf Excerpt of go-cart.io CSV spreadsheet template for the
      ``Conterminous United States'' map.}
    This template must be edited and uploaded by users to generate a
    cartogram of their own data set. The ``Population'' column is only
    for reference, while users must overwrite the third and fourth
    columns with their numeric and color data, respectively, for each
    region.}
  \label{fig:gocart_template}
\end{figure}

Secondly, go-cart.io requires the numeric and color data for each
region in the uploaded spreadsheet to be organized in a very
particular way.  Fig~\ref{fig:gocart_template} shows a spreadsheet
template for the ``Conterminous United States'' map that participants
had to edit, save, and reupload to generate their cartogram using the
spreadsheet upload feature.  In order for go-cart.io to recognize
their data, participants were required to delete the third and fourth
columns and replace them with the numeric and color data,
respectively, given during the generation task.  If participants
replaced the ``Population'' column instead, or inserted another column
before or after the ``Colour'' column, their data were either silently
ignored, or a nondescript error message ``There was a problem reading
your CSV file'' was displayed.  No participant was able to
successfully complete the go-cart.io task using the spreadsheet upload
feature without referencing the tutorial, and most made several
attempts while referencing the tutorial before they were successful.

\begin{itemize}
\item
    \textbf{Recommendation 2:}
    Web-based cartogram generation tools should attempt to automatically determine which columns in the uploaded spreadsheet contain the numeric information for target areas (e.g., population) and other visual variables (e.g., color) for each region.
\item
    \textbf{Recommendation 3:}
    If the cartogram tool is unable to parse the uploaded spreadsheet, it should provide a descriptive error message with hints about possible fixes (e.g., ``It looks like you have uploaded a spreadsheet for Austria, even though you have selected a map of the United States.'').
\end{itemize}

Finally, although the spreadsheet upload feature is the preferred
data-entry method for go-cart.io, the pop-up editing interface [shown
in panel (b) of Fig~\ref{fig:gocart-interface}] remains an
important alternative data-entry method.  Several participants used
this interface after they were unable to successfully use the
spreadsheet upload feature.  However, the pop-up editing interface
suffers from several usability challenges.  While it is styled to look
like a spreadsheet, the interface does not support basic spreadsheet
data-entry methods.  Entire columns cannot be copied and pasted, only
individual cells.  Additionally, the implementation of the color
column using the HTML color input element means that color codes
cannot be pasted at all on some browsers.

\begin{itemize}
\item
    \textbf{Recommendation 4:}
    Web-based cartogram generation tools should include a pop-up
    editing interface as an alternative to uploading a
    spreadsheet. The interface should support copying and pasting
    entire columns to and from the clipboard.
\end{itemize}

\subsubsection*{Interactive analysis tools}

We hypothesized that participants would rely on the interactive
analysis tools (infotip, linked brushing, and morphing animations)
provided by go-cart.io because
Duncan et~al.~\cite{duncan_task-based_2021} showed that these tools
improve performance during certain cartogram tasks.  However, this
hypothesis (H2) was rejected.  Multiple factors may account
for participants ignoring the interactive features.  First, many of
the analysis tasks were relatively simple.
Duncan et~al.~\cite{duncan_task-based_2021} found that interactive
analysis tools are not beneficial for simple tasks; thus, participants
in this experiment may have avoided using these tools for some tasks
because they were unnecessary.

Secondly, while the go-cart.io website itself provides interactive
analysis tools, the SVG images that it produces for export are static
and do not support any interactivity.  Because these SVG images were
used to generate the figure for the go-cart.io analysis tasks during
the experiment, the figure did not support interactivity either.  To
make use of the interactive analysis tools provided by go-cart.io,
participants had to switch back to the go-cart.io tab in their browser
during the experiment.  Thus, we believe that participants were likely
disinclined to use go-cart.io's interactive analysis tools because
they were not readily accessible.

\begin{itemize}
\item
    \textbf{Recommendation 5:}
    Web-based cartogram generation tools should include interactive
    features for analysis (e.g., infotips).  The generation tools
    should support embedding cartograms with interactive features on
    other websites.
\end{itemize}

\subsubsection*{User interface layout}

While participants complimented the aesthetics of go-cart.io's user
interface, many complained that the generation tool interface was
unintuitive.  Indeed, participants' reliance on the tutorial during
the generation tasks was much higher for go-cart.io than for fBlog.
Having good written documentation has been shown to improve the
perceived usability of a software system~\cite{davies_gis_1994}, and
most participants rated the go-cart.io tutorial as helpful or very
helpful.  However, as a web-based tool, go-cart.io cannot expect users
to read an extensive written tutorial before using the tool.  If a
user gets an initial impression that a web-based tool is difficult to
use or unintuitive, they are likely to abandon it quickly.

\begin{figure}[!ht]
  \includegraphics[width=\linewidth]{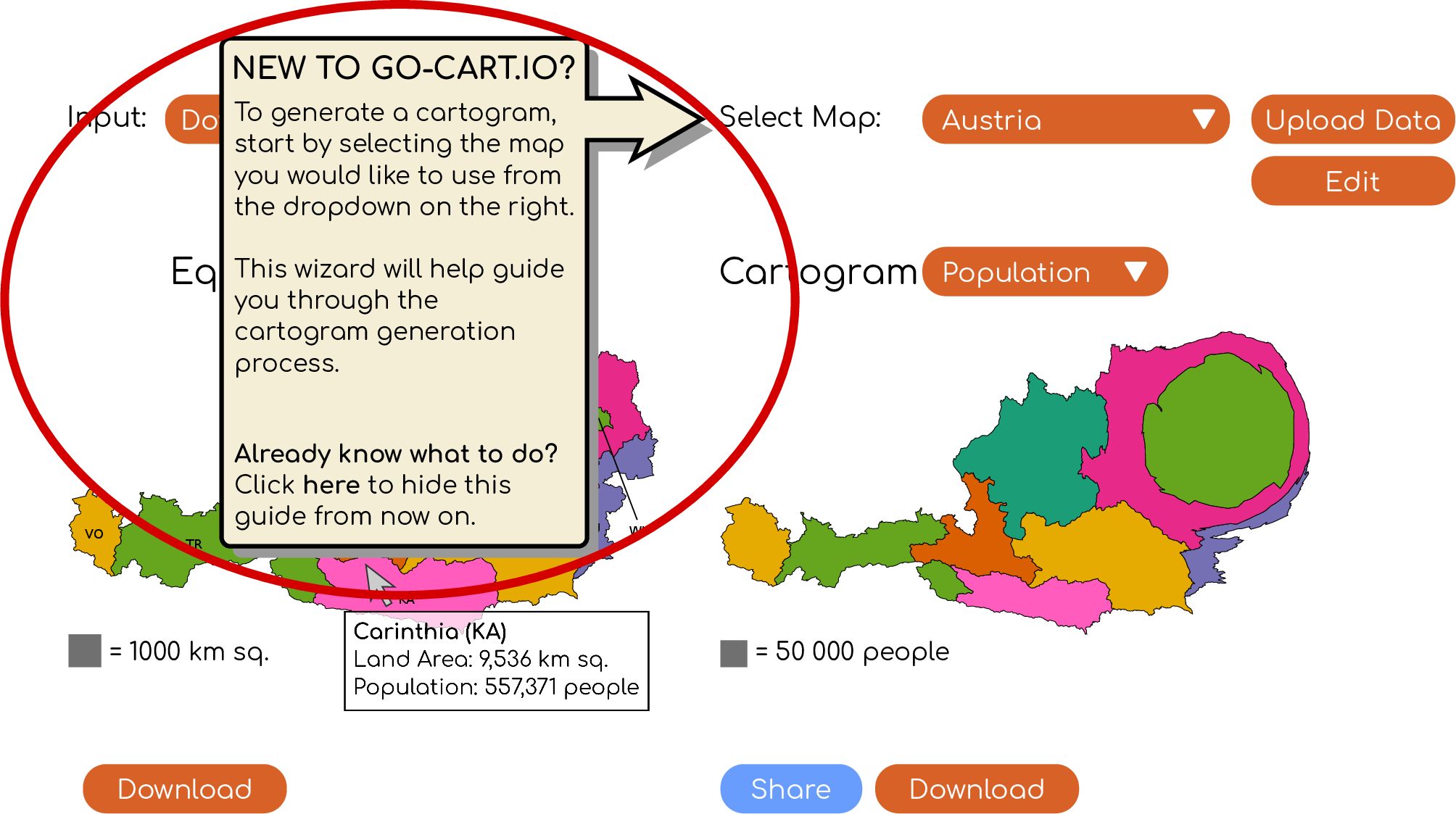}
  \caption{
    {\bf Proposed appearance of tutorial overlay.}
    Screenshot of the go-cart.io user interface with a proposed
    tutorial overlay (circled in red) to improve usability.}
  \label{fig:gocart-interface-with-tutorial}
\end{figure}

\begin{itemize}
\item
  \textbf{Recommendation 6:}
  Web-based cartogram generation tools should implement a tutorial
  overlay that guides users through the cartogram generation process
  without requiring users to reference a separate written tutorial.
  Fig~\ref{fig:gocart-interface-with-tutorial} depicts a proposed
  design for this overlay on go-cart.io.
\end{itemize}

\section*{Conclusions}

Although cartograms are becoming increasingly popular for visualizing
geospatial data, cartogram generation software has historically
suffered from poor usability, similar to many other GIS software.
While Shi, Duncan, and Gastner~\cite{tingsheng_span_2019} designed
go-cart.io with the explicit goal of creating an easy-to-use web-based
cartogram generation tool, the results of our experiment show that
go-cart.io has poor usability in key areas such as data entry.  The
purpose of this study was to make concrete recommendations for future
designs of web-based cartogram generation tools based on the usability
concerns raised by participants.

We believe that the user-centered evaluation and comparative analysis of two web-based cartogram tools (go-cart.io and fBlog) offers practical insights by focusing on real-world usage. However, we acknowledge the limitations of our study. It primarily addressed usability and user perception, omitting in-depth technical assessments of the cartogram generation process and the quality of the generated cartograms. Additionally, a broader sample size and greater participant diversity could have enhanced the generalizability of our findings. Moreover, we could have expanded the comparative analysis to include cartogram generation tools that were not in English or not web-based.

Nonetheless, it is safe to conclude that both tools that were investigated in this study (go-cart.io and fBlog) would have benefited from applying the standard guidelines of user-centered design early on in their design~\cite{lewis_task-centered_1993}. We also believe that this study reveals general usability issues that are applicable to web-based cartogram generation. We hope that our
recommendations will advance the usability of cartogram software so that
more users will be able to produce cartograms as an alternative to
traditional types of thematic maps (e.g., proportional-symbol or
dot-density maps). 
We have informed the go-cart.io developers about our recommendations.
Implementing and evaluating the effect of these
recommended practices is a source of future work.

\section*{Acknowledgments}

The authors would like to acknowledge Venkatkrishna Karumanchi’s
assistance in supervising the experiment participants. We would like
to thank Editage (www.editage.com) for English language editing and
Korneel van den Broek for permission to reproduce information and
images on the fBlog website. This project is supported by the Ministry of Education, Singapore, under its Academic Research Fund Tier 2 (EP2) programme (Award No. 653 MOE-T2EP20221-0007). Any opinions, findings and conclusions or recommendations expressed in this material are those of the author and do not reflect the views of the Ministry of Education, Singapore.

\section*{Disclosure statement}

The authors report there are no competing interests to declare.

\section*{Supporting information}

\paragraph*{S1 File.}
\label{supp:fblog_tutorial}
{\bf Tutorial of fBlog online cartogram tool.} Reprinted from~\cite{van_den_broek_online_2012} under a CC BY license, with permission from Korneel van den Broek, original copyright 2012.  Downloaded on 22 January 2022.

\paragraph*{S2 File.}
\label{supp:gocart_tutorial}
{\bf Tutorial of go-cart.io.} Downloaded from \url{https://go-cart.io/tutorial} on 22 January 2022.

\paragraph*{S3 File.}
\label{supp:analysis tasks}
{\bf Analysis tasks performed by participants during the experiment.}
Each analysis task is presented alongside the figure participants would have seen while completing the task if they completed the corresponding generation task with no errors.
Images for tasks involving fBlog were reprinted from~\cite{van_den_broek_online_2012} under a CC BY license, with permission from Korneel van den Broek, original copyright 2012.

\paragraph*{S4 File.}
\label{supp:list_of_maps}
{\bf List of maps preinstalled in go-cart.io.} This information was retrieved from \url{https://go-cart.io/cartogram} on October 28th, 2023.

\paragraph*{S1 Video.}
\label{supp:intro_video}
{\bf Introductory video shown to participant.}  All participants watched this 4-minute video before the cartogram generation and analysis tasks.

\bigskip
\noindent
The data and analysis scripts that support the findings of this study are available at \url{https://figshare.com/s/592d9e8dcf5aa933d800}.

\bibliography{go-cart-usability}

\end{document}